\documentclass[12pt,preprint]{aastex}

\def\Teff{T_{\rm eff}}
\def\teff{T_{\rm eff}}
\def\wig<{\mathrel{\hbox{\hbox to 0pt{%
          \lower.5ex\hbox{$\sim$}\hss}\raise.4ex\hbox{$<$}}}}
\def\wig>{\mathrel{\hbox{\hbox to 0pt{%
          \lower.5ex\hbox{$\sim$}\hss}\raise.4ex\hbox{$>$}}}}

\slugcomment{Accepted for publication in ApJ, July 2 2010}

\shorttitle{Properties of the T8.5 Dwarf Wolf 940 B}
\shortauthors{Leggett et al.}

\begin{document}

%% LaTeX will automatically break titles if they run longer than
%% one line. However, you may use \\ to force a line break if
%% you desire.

\title{ Properties of the T8.5 Dwarf Wolf 940 B}
% : \\ }

%% Use \author, \affil, and the \and command to format
%% author and affiliation information.
%% Note that \email has replaced the old \authoremail command
%% from AASTeX v4.0. You can use \email to mark an email address
%% anywhere in the paper, not just in the front matter.
%% As in the title, you can use \\ to force line breaks.

\author{S. K. Leggett\altaffilmark{1}}
\email{sleggett@gemini.edu}
\author{D. Saumon\altaffilmark{2}}
\author{Ben Burningham\altaffilmark{3}}
\author{Michael C. Cushing\altaffilmark{4}}
\author{M. S. Marley\altaffilmark{5}}
\and
\author{D. J. Pinfield\altaffilmark{3}}

\altaffiltext{1}{Gemini Observatory, Northern Operations Center, 670
  N. A'ohoku Place, Hilo, HI 96720} 
\altaffiltext{2}{Los Alamos National Laboratory, PO Box 1663, MS F663, Los Alamos, NM 87545}
\altaffiltext{3}{Centre for Astrophysics Research, Science and Technology Research Institute, University of Hertfordshire, Hatfield AL10 9AB}
\altaffiltext{4}{JPL, Department of Astrophysics, 4800 Oak Grove Drive, Pasadena, CA 91109}
\altaffiltext{5}{NASA Ames Research Center, Mail Stop 245-3, Moffett
  Field, CA 94035}

%% Mark off your abstract in the ``abstract'' environment. In the manuscript
%% style, abstract will output a Received/Accepted line after the
%% title and affiliation information. No date will appear since the author
%% does not have this information. The dates will be filled in by the
%% editorial office after submission.

\begin{abstract} We present 7.5--14.2 $\mu$m low-resolution spectroscopy, 
obtained with the $Spitzer$ Infrared Spectrograph, of the T8.5 dwarf 
Wolf 940 B, which is a companion to an M4 dwarf with a projected separation of 400 AU.  
We combine these data with previously published 
near-infrared spectroscopy and mid-infrared photometry, to produce the 
spectral energy distribution for the very 
low-temperature T dwarf. We use atmospheric models to derive the bolometric correction
and obtain a luminosity of $\log L/L_{\odot} = -6.01 \pm 0.05$ (the observed spectra make up 47\% of the total flux). Evolutionary models 
are used with the luminosity to
constrain the values of effective temperature ($T_{\rm eff}$) and surface gravity, and
hence mass and age for the T dwarf. 
We ensure that the spectral models used to determine the bolometric correction have
$T_{\rm eff}$ and gravity values consistent with the luminosity-implied values.
We further restrict the allowed range of $T_{\rm eff}$ and gravity using age constraints
implied by the M dwarf primary, and refine the physical properties of the T dwarf by
comparison of the observed and modelled spectroscopy and photometry. 
This comparison indicates that Wolf 940 B has a metallicity within $\sim$0.2 dex of solar, as more extreme values give poor fits to the data
--- lower metallicity produces a poor fit at  $\lambda > 2~\mu$m while higher metallicity produces a poor fit at  $\lambda < 2~\mu$m.
This is consistent with the independently derived value of [m/H]$=+0.24\pm0.09$ for the primary star, using the Johnson \& Apps (2008) $M_K$:$V-K$ relationship.
We find that the T dwarf atmosphere is undergoing vigorous mixing, with an eddy diffusion coefficient $K_{zz}$ of $10^4$ to $10^6\,$cm$^2\,$s$^{-1}$.
We  derive an effective temperature of 585 K to 625 K,  and 
surface gravity $\log \ g = 4.83$ to 5.22 (cm s$^{-2}$), for an age range 
of 3 Gyr to 10 Gyr, as implied by the  kinematic and H$\alpha$ properties of the M dwarf 
primary. Gravity and temperature are correlated such that the lower gravity corresponds to 
the lower temperature and younger age for the system, and the higher values to the higher temperature and older age. The mass of the T dwarf
is 24 M$_{\rm Jupiter}$  to 45 M$_{\rm Jupiter}$ for the younger to older age limit.

\end{abstract}

%% Keywords should appear after the \end{abstract} command. The uncommented
%% example has been keyed in ApJ style. See the instructions to authors
%% for the journal to which you are submitting your paper to determine
%% what keyword punctuation is appropriate.


%% From the front matter, we move on to the body of the paper.
%% In the first two sections, notice the use of the natbib \citep
%% and \citet commands to identify citations.  The citations are
%% tied to the reference list via symbolic KEYs. The KEY corresponds
%% to the KEY in the \bibitem in the reference list below. We have
%% chosen the first three characters of the first author's name plus
%% the last two numeral of the year of publication as our KEY for
%% each reference.

\section{Introduction}

The study of brown dwarfs, stellar-like objects with a mass below that 
required for the onset of hydrogen fusion (e.g. Burrows et al. 2001), has advanced 
dramatically in the last decade. Primarily this has been due to the 
discovery of a significant population of brown dwarfs by
far-red and near-infrared surveys: the Sloan Digital Sky 
Survey (SDSS, York et al. 2000), the Two Micron All Sky Survey (2MASS, 
Skrutskie et al.  2006), the UK Infrared Telescope (UKIRT) Infrared Deep 
Sky Survey (UKIDSS, Lawrence et al. 2007) and the Canada France Hawaii 
Telescope's Brown Dwarf Survey (CFBDS; Delorme et al. 2008b). The latest 
spectral type and coolest dwarfs currently known are classified as T 
dwarfs. At the time of writing there are eight objects known with type 
later than T8: ULAS J003402.77-005206.7 (T9, hereafter ULAS 0034-00, Warren 
et al. 2007), CFBDS J005910.90-011401.3 (T9, Delorme et al. 2008a), ULAS 
J123828.51+095351.3 and ULAS J133553.45+113005.2 (T8.5 and T9, Burningham et al. 
2008), Wolf 940 B (T8.5, Burningham et al. 2009, hereafter B09), 
ULAS J130217.21+130851.2 
(T8.5, Burningham et al. 2010), Ross 458 C (T8.5, Goldman et al. 2010) and UGPS J072227.51-054031.2
(T10, Lucas et al. 2010).  These extreme objects have very low effective temperatures 
($T_{\rm eff}$) of 500 -- 600~K (e.g. Leggett et al. 2010), approaching the temperature
where water clouds are expected to form in the atmosphere (e.g. Burrows et al. 2003).

The T dwarfs that have stellar companions are of
particular interest as the primary can provide distance, metallicity and age constraints
for the brown dwarf. There are ten known examples of such systems at this time. In order
of increasing T dwarf sub-class they are:
Gl 337 CD (T0, T0 $+$ K1V $+$ G9V; Wilson et al. 2001, Burgasser et al. 2005);
Epsilon Indi Bab (T1, T6 $+$ K5V; Scholz et al. 2003, McCaughrean et al. 2004);
HN Peg B (T2.5 $+$ G0V;  Luhman et al. 2007, Leggett et al. 2008);
SCR 1845-6357 B (T6 $+$ M8.5; Biller et al. 2006, Kasper et al. 2007);
Gl 229 B (T7p + M1V; Nakajima et al. 1995; Leggett et al. 2002);
HD 3651 B (T7.5 $+$ K0V; Mugrauer et al. 2006, Burgasser 2007, 
Liu et al. 2007, Luhman et al. 2007);
Gl 570 D (T7.5 $+$ M3V $+$ M1.5V $+$ K4V; Burgasser et al. 2000, Saumon et al. 2006, Geballe et al. 2009); 
Wolf 940 B (T8.5 $+$ M4V, B09);
Ross 458 C (T8.5 $+$ M7V $+$ M0.5; Goldman et al. 2010).
The degree of accuracy that can be obtained in measurements of the fundamental parameters of such
T dwarfs is illustrated by the Geballe et al. (2009) study of the T7.5 dwarf Gl 570D.
The Gl 570 system has a well determined distance and metallicity, and the K4V
primary constrains the age to be between 2 and 5~Gyr (Geballe et al. 2001).
Comparison of the observed luminosity and spectral properties of the T dwarf to
atmospheric and evolutionary models constrains $T_{\rm eff}$ to be 800 -- 820~K
and the gravity to be  $\log \ g = 5.09$ -- 5.23. These values constrain
the age further to  3 to 5~Gyr, and the mass of the T dwarf to be between 
38 and 47 M$_{\rm Jupiter}$. 

This paper presents an analysis of the spectral energy distribution (SED) of 
the T8.5 dwarf Wolf 940 B. The dwarf was discovered in the Large Area 
Survey component of the UKIDSS, and identified as a companion to the M4V 
star Wolf 940 (also known as LHS 3708 and GJ 1263), by B09.
B09 present near-infrared spectroscopy and photometry, as 
well as $L^{\prime}$ photometry, for the T dwarf. They use these data to 
derive the luminosity of the dwarf, and hence 
$T_{\rm eff}$ and surface gravity. Leggett et al. (2010) give 
$Spitzer$ Infrared Array Camera (IRAC, Fazio et al. 2004) [3.6], [4.5], 
[5.8] and [8.0] photometry for the T dwarf and use the infrared colors 
to support and further constrain the atmospheric properties determined by 
B09. Here we report new mid-infrared spectroscopy for Wolf 
940 B, obtained with the $Spitzer$ Infrared Spectrograph (IRS, Houck et 
al. 2004). The improved accuracy of the luminosity we derive, and the additional spectroscopy, enables us to 
refine the B09 results. Wolf 940 B is currently the
coolest companion to a star with both near- and mid-infrared photometric
and spectroscopic data.
A more complete energy distribution for such a cool object, which has 
other basic data provided by its companion (distance, age and metallicity), provides a good test of the 
model atmospheres for these almost-planetary objects. 

In \S 2 we describe the Wolf 940 system. The IRS data are presented in 
\S 3. In \S 4 we use atmospheric and evolutionary models to 
calculate an accurate luminosity for Wolf 940 B, and constrain its 
fundamental properties. Our conclusions are given in \S 5.

\section{The Wolf 940 System}

%\subsection{Basic Data}

The Wolf 940 binary consists of a widely separated M4 and T8.5 dwarf pair
(Reid et al. 1995, B09).
Table 1 lists the astrometric and photometric properties of the system.
The primary is a high proper-motion M4 dwarf, identified in 
historic proper motion surveys  as Wolf 940 (and LHS 3708, GJ 1263, amongst other 
identifiers). 

B09 use the colors of the primary and the $M_K$:$V-K$ relationships presented by 
Bonfils et al. (2005) to derive a metallicity [m/H]$=-0.06\pm0.20$ 
for the system. However Leggett et al. (2010) show that the $H-K$ and 
$H-$[4.5] colors of the secondary imply that the system has solar or 
slightly super-solar metallicity. Here we use the Johnson \& Apps (2008)
significant revision of the Bonfils et al. calibration, together with the parallax, $V$ and $K$ values given in Table 1, to derive a metallicity of [m/H]$=+0.24\pm0.09$ 
for the system. We discuss metallicity further in \S 4.6.

The measured radial velocity, combined with the parallax
and proper motion, give UVW space motions of $35$, $-49$ and $-26$ kms$^{-1}$
(B09 and references therein), typical of old disk objects.
B09 match the combination of the kinematics and the Bonfils et al. metallicity value
to the 3 -- 5 Gyr-old disk population in the Galaxy population synthesis model by 
Robin et al. (2003).
In this paper we constrain the age of the system using the 
Geneva-Copenhagen survey of age, kinematics and metallicity of a large number 
of stars in the solar neighbourhood (Holmberg et al. 2009). Based on this work, the UVW velocities and revised metallicity of the Wolf 940 system implies an
age of between 3 Gyr and 10 Gyr. We adopt this age range,
taking a more conservative approach than B09.

Wolf 940 A shows the H$\alpha$ line in absorption, with an equivalent 
width of $-0.26$ \AA, typical for an M4 dwarf (Gizis et al. 2002, their Figure 6). 
The evolution of chromospheric activity in low mass stars is a complex 
process, and its relationship to the strength of the H$\alpha$ line is not well
understood (e.g. Gizis et al. 2002, West et al. 2008, Walkowicz \& Hawley 2009).
While it is clear that H$\alpha$ in emission implies a highly active 
and therefore young star,
stars with H$\alpha$ in absorption may have no to moderate activity.
Given that Wolf 940 A shows the line in absorption with a very typical strength
for its type, we assume that it is no longer active and is therefore 
older than 3 Gyr, which is the lower limit on the activity lifetime for an
M4 dwarf (West et al. 2008, their Figure 10). Again we take a more conservative
approach than B09, and do not assume that the presence of H$\alpha$ 
absorption implies that the star is still somewhat active, and so
we do not put an upper limit on the age of the system based on 
chromospheric activity arguments.

%\subsection{Age}

In summary, the kinematic and H$\alpha$ properties of Wolf 940 A imply 
that the 
system is between 3 and 10 Gyr old. A rotational 
velocity for Wolf 940 A would be useful as it would allow a 
gyrochronology age to be determined. The metallicity of the system is approximately solar.

\section{Observations}

We obtained $Spitzer$ IRS spectra of Wolf 940 B via the Director's Discretionary
Time program 527. The Short-Low module was used in first order which provides 
a spectrum covering the 7.5 -- 14.5 $\mu$m range, with a resolution of $R \approx 120$.
A nearby star was used in the blue peak-up array to acquire the target  
in the 3$\farcs$7 slit. Standard staring mode was used with a ramp duration of 60s 
and 84 cycles, for an observation of duration 3.6 hours; two such observations 
were obtained. Each cycle places the target at two slit positions, about
one-third and two-thirds of the way along the slit. Thus the total on-source
exposure was 5.6 hours. Both observations were carried out on the 16th December 2008.

We used the Basic Calibrated Data (BCD) produced by the $Spitzer$ pipeline version
S18.7.0. This produces a two-dimensional long-slit spectrum at each of the nod
positions, so for each of our two observations  there are 84 files for each
nod position. The BCD files have been processed by the pipeline, which includes ramp
fitting, dark subtraction, droop correction, linearity correction, flat fielding, and wavelength calibration (see Section 5.1 of the IRS Instrument Handbook
\footnote{http://ssc.spitzer.caltech.edu/irs/irsinstrumenthandbook/49/}).

We treated the two observations as independent datasets. We used the IRAF 
data reduction tool (Tody 1993) to subtract each nod pair, and combined the 84 
subtracted images
using sigma-clipping in the {\tt imcombine} routine.
Some structure remained in the background of the combined images; we removed this 
by subtracting a smoothed version of the image, using an $11\times 11$ box.
This residual-structure subtracted image was then fed into the $Spitzer$ IRS data 
reduction package SPICE. 
\footnote{http://ssc.spitzer.caltech.edu/dataanalysistools/tools/spice/spiceusersguide/1/}
The source signal was low, and there were a large number of bad and hot pixels.

SPICE takes as input: the image file, a bad pixel mask and an uncertainty image. We used the
sigma image produced by the {\tt imcombine} routine as the uncertainty mask, and a bad pixel mask 
included with the bcd files. SPICE traces the appropriate module's slit profile over the 
input image, allowing the user to set the location and the width of the extraction window.
We extracted the positive and negative spectrum using a relatively small window of 2 pixels
in order to avoid adding noise. The final step in the SPICE application converts the signal 
to a flux density.

We combined the two negative and two positive extracted spectra manually. The signal to noise 
is quite low (S/N $\approx 4$--10), and we visually identified bad data points by comparison of the four spectra. 
Figure 1 shows the final spectrum, with the noise spectrum which is based on the deviation between the four individual spectra. We checked the flux calibration by 
synthesizing the IRAC [8.0] photometry from the spectrum, after extending it to the short
end of the filter bandpass ($\sim 6.3~\mu$m) using as a template the IRS 
second order spectrum
of a bright T8 dwarf (this extension contributed 17\% of the total flux through the filter). This check indicated that the flux calibration is good to 5\%.

\section{Analysis}

Our analysis is based on a method we developed that uses synthetic
spectra to determine a bolometric correction that is consistent with
the luminosity obtained from evolution models.  Because the effective
temperatures and gravities obtained in this fashion are obtained from the
{\it integrated} observed flux and not by fitting the observed spectrum, it
is  quite robust and is not sensitive to the known (and unknown) systematic
biases that arise when fitting model spectra to the observed spectrum.

The solution for the $\teff$ and gravity is further refined by direct
comparison of the corresponding synthetic spectra with the spectroscopic
and photometric data, as well as age constraints. In this section, we
first obtain $\teff$ and the gravity using solar metallicity models.
The mid-infrared photometry and spectrum allow an exploration of
non-equilibrium chemistry driven by vertical transport and we estimate
the mixing time scale.  Finally, we perform the analysis again with
non-solar metallicity models and by considering the uncertainty in
$L_{\rm bol}$.

\subsection{Effective Temperature and Gravity}

Our analysis of Wolf 940 B is based on the atmospheric models and non-equilibrium
chemistry scheme described in Saumon et al. (2006, 2007) and the evolution sequences 
of Saumon \& Marley (2008). We use cloudless atmospheres and evolution as appropriate 
for a late T dwarf, and initially assume solar metallicity.  The observed 1.05 -- 2.43 
~$\mu$m spectrum (B09) and 7.5 -- 14.2~$\mu$m spectrum (this work) are integrated to
give an observed flux at Earth of $9.22 \pm 0.24 \times 10^{-14}\,$ erg$\,$s$^{-1}\,$cm$^{-2}$. 
The IRS spectrum contributes 39\% of this value.  Following the method described in
Geballe et al. (2001) and Saumon et al. (2006, 2007), we derive a family of $(\teff,g)$ values
that provide $L_{\rm bol}$ values that are consistent with the bolometric correction
from the model spectra and the evolution.  That is, the  $(\teff,g)$ of the spectral model used
to calculate the correction from observed to total flux,  must be consistent with
the  $(\teff,g)$ implied by the evolutionary models for that total luminosity.
These $(\teff,g)$ solutions are shown in Figure 2 
and Table 2, and correspond approximately to a constant $L_{\rm bol}$ curve in 
the $(\teff,g)$ plane.  The bolometric correction amounts to $\sim53$\% of the luminosity, 
and the derived luminosity is $\log L_{\rm bol}/L_{\odot} = -6.009$ (for $T_{\rm eff}=600\,$K, see below and Table 2).
Combining the uncertainty in the flux calibration of the spectrum (using a Monte Carlo
sampling of a Gaussian distribution of the calibration uncertainty for each segment of the
spectrum) and in the distance, the uncertainty in $L_{\rm bol}$ is $\pm 0.047\,$dex.

Figure 2 shows evolutionary sequences in a $T_{\rm eff}$ against $\log \ 
g$ plot; the allowed set of values for Wolf 940 B obtained with the method
outlined above is indicated by the sequence of solid dots.
The $\teff$ is thus constrained to be in the range of 500 to 640$\,$K.
This range can be narrowed from a knowledge of the age of the system, or by comparison of the data to model spectra. The $(\teff,g)$ of the models used to generate the synthetic spectra for such a data comparison must fall between the dashed red lines of
Figure 2, which corresponds to $\log L_{\rm
bol}/L_{\odot}=-6.009\pm0.047$.  Figure 2 also demonstrates how each ($T_{\rm eff}$, $\log \ g$)  
pair determines the dwarf's mass and age. Table 2 gives the set of 
allowed parameters ($T_{\rm eff}$, $\log \ g$, $\log L_{\rm bol}/L_{\odot}$, mass, radius and age) in steps of  $\Delta T_{\rm eff} = 25\,$K for ages ranging from 0.4 to 10 Gyr 
and for solar metallicity. The impact of varying metallicity is 
discussed below in \S 4.6. 

\subsection{Comparison with the Near-Infrared Spectrum}

Having constrained the allowed $(\teff,g)$ parameter range, we
compare the corresponding synthetic spectra with the data.
Figure 3 shows such a comparison
for the solutions with ages of 1, 5.5 and 10 Gyr, i.e. the $(T_{\rm eff}$, $\log \ g)$ 
solutions (575, 4.722), (600, 4.989) and (625, 5.221) (see Table 2). 
Note that the scaling of the model fluxes to the data is not adjustable: it is 
fixed by the measured distance to the Wolf 940 system and the radius given
in Table 2, determined 
by evolutionary models from the values of ($T_{\rm eff}$, $\log g$).
Here we have assumed
solar metallicity and an eddy diffusion coefficient $K_{zz}=10^6\,$cm$^2\,$s$^{-1}$.  
The eddy diffusion coefficient parameterizes the time scale of vertical mixing
in the upper radiative region of the atmosphere, that drives the chemistry of
carbon and nitrogen out of equilibrium (Saumon et al. 2006, 2007; Hubeny \& Burrows 2007). In this analysis, we have considered $K_{\rm zz}=0$
(chemical equilibrium), $10^2$, $10^4$ and $10^6\,$cm$^2\,$s$^{-1}$.  
T dwarfs typically have values of $K_{zz}$ between $10^4$ and $10^6\,$cm$^2\,$s$^{-1}$
(e.g. Saumon et al. 2007, Leggett et al. 2009, Stephens et al. 2009); the effect of
varying $K_{zz}$ is discussed further below.
In the deeper convection zone, the mixing time scale
is determined from the mixing length formulation of convection. 

As shown by Figure 3, and as we have found for other late T dwarfs, the near-infrared spectra of models
constrained with our method are very nearly identical (Saumon et al. 2007, Geballe et al. 2009).  Since these models have
very nearly the same $L_{\rm bol}$, it is a fairly accurate statement that the
near infrared spectra of low--$\Teff$ models are independent of gravity {\it at fixed}
$L_{\rm bol}$, for a given metallicity. This is because the effect of increasing the gravity is partially canceled by the accompanying increase in effective temperature.
Also note that the effect of vertical mixing is not significant in the near-infrared at these $T_{\rm eff}$ (e.g. Saumon et al. 2006, and \S 4.4 below).

The match to the near-infrared spectrum
is poor, with the modelled $J$ and $H$ peaks high and the $K$ low (Figure 3, top panel). This has been 
noticed before and is likely due to the known incompleteness of the 
molecular opacity line lists in this region (e.g. Saumon \& Marley 2008).

\subsection{Comparison with the Mid-Infrared Spectrum}

In contrast to the near-infrared region, differences are seen between the mid-infrared synthetic spectra generated by the approximately constant-luminosity models (Figure 3, lower panel). Calculation of the reduced $\chi^2$
between the models and the observed mid-infrared 
spectrum shows that the best match is the $\teff=600\,$K model ($\chi^2=3.64$), which is 
better than the 575$\,$K ($\chi^2=4.93$) and 625$\,$K ($\chi^2=4.29$) solutions.  The 
$\teff=550\,$K solution is clearly excluded with $\chi^2=10.7$. 

Vertical transport in the convection zone drives the chemistry of nitrogen away
from equilibrium and results in a typical depletion of NH$_3$ of a factor of 8 -- 10
(e.g. Saumon et al. 2006). This increases the overall flux level in the 9 -- 15$\,\mu$m region where absorption by NH$_3$ dominates.  A comparison of the observed 
spectrum and synthetic spectra computed with equilibrium chemistry
gives large  $\chi^2$ values --- from 6.9 (550$\,$K solution) to 11.7 (625$\,$K solution).
Those higher $\chi^2$ values arise from the inability of the equilibrium models
to match the overall flux level of the IRS spectrum for $\lambda>9\,\mu$m (Saumon et al. 2006, see also further discussion below).
We can conclude that there is convincing evidence for a depleted 
abundance of NH$_3$ in the atmosphere of Wolf 940 B, which is consistent with 
findings from all of the other IRS spectra of late T dwarfs (Saumon et al. 2006, 2007;
Burgasser et al. 2008, Leggett et al. 2009).

\subsection{Comparison with 3 -- 9~$\mu$m Photometry}

Synthetic fluxes in the $L^\prime$ and IRAC band passes were obtained by integrating the model
spectra over the filter band passes, and these are shown in Figure 3.
The IRAC [3.6] photometry is not well matched by any of the models. This 
is a known model deficiency; Leggett et al. (2009, 2010) show that the 
calculated [3.6] fluxes are systematically low by 20 -- 50\%. It is not 
clear if this is related to missing opacities in the near-infrared 
(which may lead to too high near-infrared fluxes and too low 
mid-infrared fluxes), or if it is due to distortions of the 
pressure-temperature structure of our non-equilibrium models.

The reduced $\chi^2$ between the synthetic and measured photometry in these five
bands strongly favor the highest $\teff$ model at 625$\,$K as well
as a value of the eddy diffusion coefficient of $\log K_{zz} \approx 6$.  However,
the $\chi^2$ is dominated by a large contribution from the IRAC [3.6]
band pass (Figure 3).  If we ignore this flux measurement, then all three models with 
$\teff \ge 575\,$K and $\log K_{zz}=6$ match equally well and are far better
than all other parameter combinations.  For all four models from Table 2, the
spectrum computed in chemical equilibrium ($K_{zz}=0$) is by far the worst.
Thus, the photometry strongly favors models that depart from chemical equilibrium.

For a fixed metallicity, the IRAC [4.5] flux is a sensitive probe of non-equilibrium
carbon chemistry as the filter bandpass covers the fundamental band of CO centered 
at 4.65$\,\mu$m.  More vigorous vertical transport (i.e. larger
$K_{zz}$) leads to an over-abundance of CO in the upper atmosphere and a strong
band where none should be detectable at such low $\teff$ (Noll et al. 1997,
Golimowski et al. 2004, Patten et al. 2006, Geballe et al. 2009).
A low [4.5] flux is thus a hallmark of excess CO and of a non-equilibrium
CO abundance. The observed [4.5] flux for Wolf 940 B is well below what would be expected from chemical equilibrium and only non-equilibrium models can match its
value.  We find that $\log K_{zz}=5.5$ gives an excellent match and Figure 4 shows 
our best fitting non-equilibrium model spectrum, along with the corresponding equilibrium spectrum ($K_{zz}=0$). 
A diffusion coefficient of $K_{zz} = 10^{5.5}$ cm$^2\,$s$^{-1}$ corresponds to a vertical mixing time scale of $\sim 4\,$hours.
Note in Figure 4 the large difference between the non-equilibrium and equilibrium models in [4.5] flux, and in the flux level in the 9 -- 15$\,\mu$m region where absorption by NH$_3$ dominates. Finally, we note that fitting $K_{\rm zz}$ independently from $\teff$, gravity and $L_{\rm bol}$ affects $L_{\rm bol}$ by less than 0.016 dex, which is small compared to the contributions from the distance and flux calibration uncertainties of 0.047 dex. 

%The evidence in favor of non-equilibrium chemistry driven by vertical transport is
%thus: 1) the high flux level of the spectrum beyond 9$\,\mu$m and 2) the four photometric
%channels $L^\prime$ and IRAC [4.5], [5.8], and [8.0].  Fitting the [4.5] flux
%gives the value of $K_{zz}$. 

\subsection{Optimal Parameters for Solar Metallicity}

Based on our solar metallicity model atmospheres, spectra and evolution, we conclude
that Wolf 940 B is best fitted with $\teff=575-625\,$K, $\log g=4.72 - 5.22$
and $\log K_{\rm zz} \approx 5.5$.  If we apply the age constraint of 
3 -- 10$\,$Gyr determined by the primary,
then the parameter range is further restricted to $\teff=585-625\,$K and 
$\log g=4.83-5.22$. The derived parameters $\teff$, $\log g$,  $K_{\rm zz}$, mass, radius and age are summarized in Table 3. As shown in Figure 2, the values of $\teff$ and $\log g$
are correlated, thus Wolf 940 B cannot be both low-$\teff$ and high gravity.  
These ranges do not reflect the effect of the uncertainty
in the determination of $L_{\rm bol}$ of 0.047 dex, which translates to 
$\pm 16\,$K in $\teff$ (for a fixed gravity) and $\pm 0.16$ dex in gravity
(for a fixed $\teff$).

\subsection{The Effects of Varying Distance and Metallicity}

While we have been able to obtain a reasonably good fit of all the data
available (parallax, spectra and photometry), systematic differences
between the models and the near-infrared spectrum linger.  In particular, we
find that the models are much bluer in $J-K$ than the data (Figures 3 and 4). The
$K$ band modelled flux is about half of what is observed. A blue $J-K$ color in a
late T dwarf is an indicator of a high-gravity or low-metallicity object (e.g. Knapp et al. 2004).  
Given all the constraints applied in our method of analysis, we cannot change
the gravity independently of $\teff$ (except within the error bars) and we
have seen that this has no effect on the $K$ band flux (Figure 3).  We can, however,
consider variations in metallicity and of the distance, within the range allowed
by the parallax uncertainty.

We have redone the complete analysis by assuming that the luminosity is
increased by $1 \sigma$ to $\log L_{\rm bol}/L_\odot=-5.939$ (see Figure 2) and keeping
[m/H]$=$0.  For consistency, this new luminosity implies that the distance is 
increased by $1.4\sigma$ to 13.58$\,$pc.  This moves our best fitting model
of $\teff=600\,$K and $\log g=4.989$ to $\teff=625\,$K and $\log g=5.0$.
This worsens the fit slightly, but well below the uncertainties and is
thus not significant.

The analysis was performed again with model atmospheres, spectra and evolution
of different metallicities.  We find that a variation of $\Delta[m/H]=\pm0.3$
results in $\log L_{\rm bol}=\mp 0.023$, which is half of the uncertainty on
$L_{\rm bol}$ due to the photometric and parallax uncertainties. Since 
the uncertainties in $L_{\rm bol}$ are effectively negligible in terms
of spectrum fitting, for clarity of presentation we will make a comparison 
of spectra of different metallicity calculated with the best 
fitting $(\teff,\log g)=(600, 5.0)$ we obtained in the solar metallicity case.  
 
Spectra with [m/H]$=-0.3$ and $+0.3$  are shown in Figure 5. 
The optimal value of $K_{\rm zz}$ depends on [m/H].  To better show the role of [m/H]
in shaping the spectrum, we keep $K_{\rm zz}$ fixed at $10^4\,$cm$^2\,$s$^{-1}$ for the non-solar metallicity models. As expected, increasing the 
metallicity raises the modelled $K$ band flux and a very good agreement is obtained for
[m/H]$= +0.3$, but this comes at the cost of a significant increase in the modelled flux
in the $Y$, $J$ and $H$ bands. On the other hand, the spectrum with [m/H]$=-0.3$
is an excellent match to the observed $YJH$ peaks but increases the discrepancy at $K$.
Calculation of the reduced $\chi^2$ confirms the qualitative visual impression from the top
panel of Figure 5 that the [m/H]$=-0.3$ spectrum is a better fit to the near infrared
spectrum of Wolf 940 B.  The lower panel shows the opposite situation at mid-infrared
wavelengths: the fit of the spectrum improves steadily as the metallicity
increases from $-0.3$ to $+0.3$.  The mid-infrared photometry also favors higher metallicity
if we neglect the poorly fit 3 -- 4 ~$\mu$m region (where there is little flux).
Hence varying the metallicity from [m/H]$=-0.3$ to +0.3 has a seesaw
effect on the shape of the spectrum compared to the data and provides no real improvement
over the solar metallicity fit.  Since the extremes of that range lead to notably poor
fits in the near-infrared or the mid-infrared, we can reasonably conclude that
Wolf 940 B has a metallicity within $\sim$0.2 dex of solar, which is consistent with the value of [m/H]$=+0.24\pm0.09$ derived for the primary star using the Johnson \& Apps (2008) $M_K$:$V-K$ relationship (see \S 2). 

Figure 5 (lower panel) shows that the [4.5] flux is not only  sensitive
to $K_{\rm zz}$, it is also sensitive to [m/H]. The best solar metallicity solution gives
$\log K_{zz} \approx 5.5$ using this datapoint, as described above in \S4.4. If [m/H]$=+0.3$ then $\log K_{zz} \approx 4$ reproduces the observed [4.5] flux, and similarly  if [m/H]$=-0.3$ then $\log K_{zz} >7$. 
Values of $\log K_{zz} =$ 4, 6 and 7 correspond to  mixing time scales of 2.8 days, 1.6 hours and 14 minutes respectively. The convective mixing time scale is 0.6 min for comparison, meaning that $\log K_{zz} >7$ implies that mixing is almost as vigorous as convection which may not be realistic. A metallicity as low as [m/H]$=-0.3$ appears unlikely based on the very high value of $K_{\rm zz}$ required, again consistent with the
solar or slightly metal-rich solution found for the primary.

\section{Conclusions}

%%Note that we have strong evidence for non-equil chemistry and a more
%%rigorous analysis and fit than Burningham et al (2009) did.  Our results
%%are not sensibly different from theirs though. No splashy result to
%%report!

We have presented a 7.5 -- 14.2~$\mu$m low-resolution spectrum
 of the T8.5 dwarf Wolf 940 B, which is a companion to an M4 dwarf with a projected separation of 400 AU.
This spectrum complements the near-infrared spectrum and $L^{\prime}$ photometry
presented by B09, and the IRAC photometry presented by Leggett et al. (2010).
Combining all these data allows a rigorous analysis of the dwarf's SED. 
Assuming an age range of 3 -- 10 Gyr as indicated by the primary, 
evolutionary and atmospheric models show that for the T dwarf secondary 
 $\teff=585-625\,$K and $\log g=4.83-5.22$. Gravity and temperature are correlated such that the lower gravity 
corresponds to the lower $\teff$ and younger age for the system, and the higher value to
the higher $\teff$ and older age. The mass of the T dwarf
is 24 M$_{\rm Jupiter}$  to 45 M$_{\rm Jupiter}$ for the younger to older age limits.
An age younger than 1 Gyr is excluded by the analysis of the T dwarf spectrum alone.

These temperatures and gravities are slightly higher than those derived in B09.
Our luminosity, which benefits from the inclusion of the IRS spectrum, is slightly (but
not significantly) higher than determined by B09. Our higher temperature and gravity range principally
arises from our more conservative approach to the age of the primary, for which we adopt 3 -- 10 Gyr as opposed
to B09's 3 -- 5 Gyr.

The IRS spectrum demonstrates that vertical mixing is important in the radiative zone of the atmosphere of Wolf 940 B,  as is the case for many, if not all, T dwarfs. The [4.5] photometric datapoint constrains $\log K_{zz}$ to 5.5 for solar metallicity. 
If the metallicity is higher then $K_{zz}$ is lower, and vice versa.  Comparison of the
synthetic SED to the observational data shows that
Wolf 940 B has a metallicity within $\sim$0.2 dex of solar.  
More extreme values give poor fits to the data
--- lower metallicity produces a poor fit at  $\lambda > 2~\mu$m while higher metallicity produces a poor fit at  $\lambda < 2~\mu$m.
This is consistent with the value of [m/H]$=+0.24\pm0.09$ derived here for the primary star using the Johnson \& Apps (2008) $M_K$:$V-K$ relationship.

It is very useful to find T dwarf companions to main sequence stars, given that the brown dwarfs are cooling and hence mass
is difficult to constrain without knowing age. We look forward to more examples like Wolf 940 B (B09) and Ross 458 C (Goldman et al. 2010)
being found, as the UKIDSS sky area increases.

%% If you wish to include an acknowledgments section in your paper,
%% separate it off from the body of the text using the \acknowledgments
%% command.

%% Included in this acknowledgments section are examples of the
%% AASTeX hypertext markup commands. Use \url without the optional [HREF]
%% argument when you want to print the url directly in the text. Otherwise,
%% use either \url or \anchor, with the HREF as the first argument and the
%% text to be printed in the second.

\acknowledgments

This work is based  on observations made with the {\it Spitzer Space Telescope}, which is operated by the Jet Propulsion Laboratory, California Institute of Technology under a contract with NASA. Support for this work was provided by NASA through an award issued by JPL/Caltech. Support for this work was also provided by the Spitzer Space Telescope Theoretical Research Program, through NASA.
SKL's research is supported by the Gemini Observatory, which is operated by the Association of Universities for Research in Astronomy, Inc., on behalf of the international Gemini partnership of Argentina, Australia, Brazil, Canada, Chile, the United Kingdom, and the United States of America.

\clearpage

%% Use the figure environment and \plotone or \plottwo to include 
%% figures and captions in your electronic submission.

\begin{figure}
\includegraphics[height=.85\textheight,angle=0]{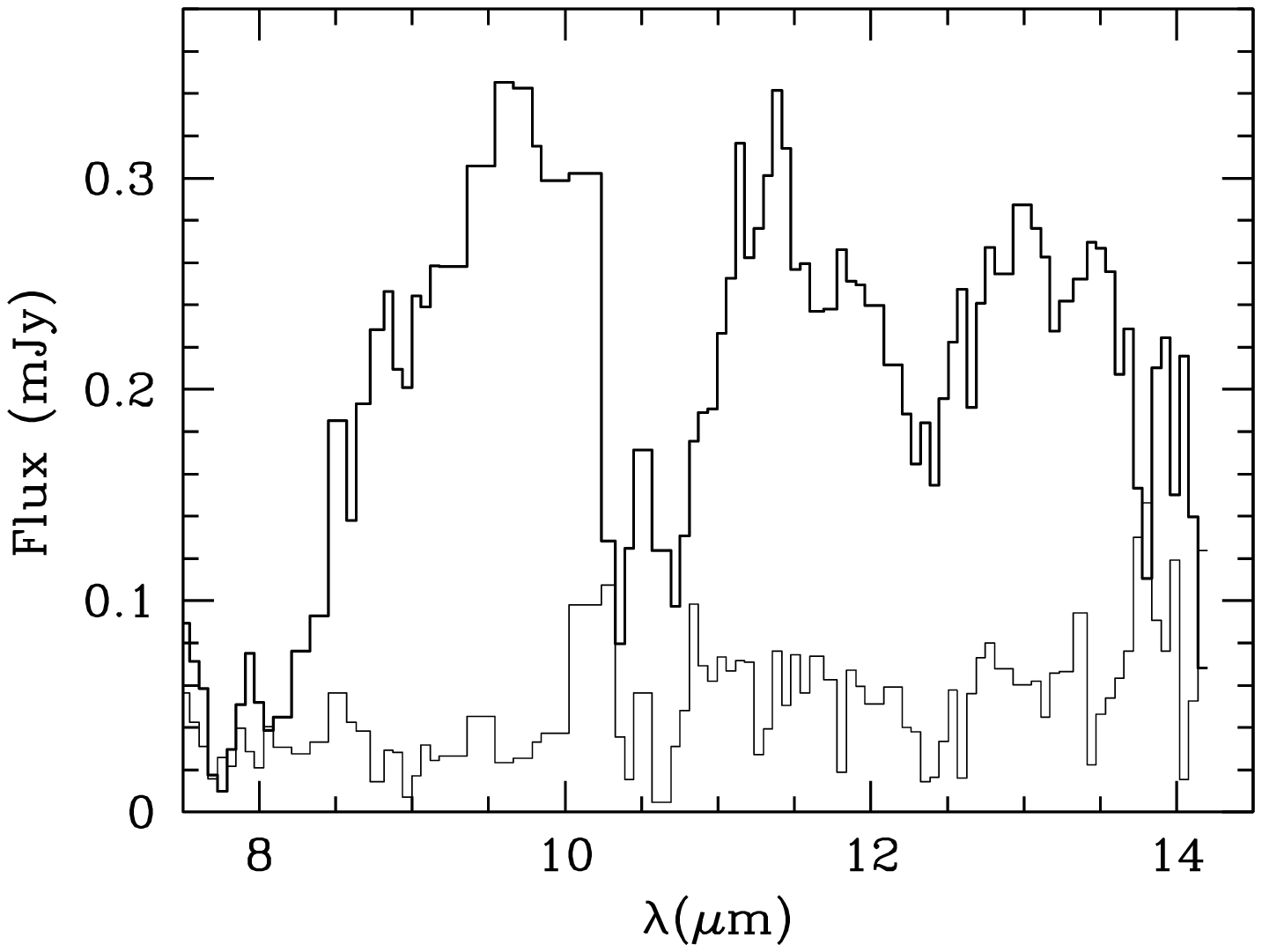}
\caption{IRS spectrum of Wolf 940 B, the noise spectrum is shown in the lower region of the plot.}
\end{figure}

\clearpage 

\begin{figure}
\includegraphics[height=.8\textheight]{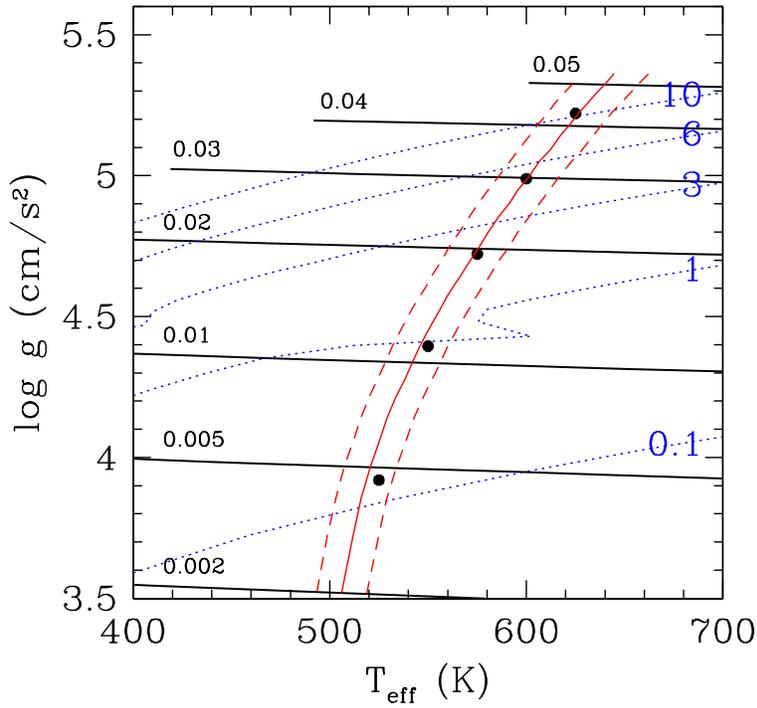}
\caption{Evolution of cloudless brown dwarfs and constrained values of $(\teff,g)$
         for Wolf 940 B. Brown dwarf cooling tracks (solid black lines) are labeled with
         the mass in M$_\odot$.  Isochrones (blue dotted curves) are labeled with the
         age in Gyr. Solid dots are solutions for Wolf 940 B for $\teff=525$ to 625$\,$K in
         steps of 25$\,$K (see Table 2).  A constant luminosity curve 
         ($\log L_{\rm bol}/L_{\odot}=-6.009$) is plotted to guide the
         eye (solid red curve), as well as its uncertainty ($\pm 0.047 dex$, dashed red curves).
         We estimate the age of the Wolf 940 B system to be between 3 and 10$\,$Gyr.
         The spur on the 1$\,$Gyr isochrone is caused by deuterium burning.
          }
\end{figure}

\clearpage 

\begin{figure}
\includegraphics[height=.8\textheight,angle=0]{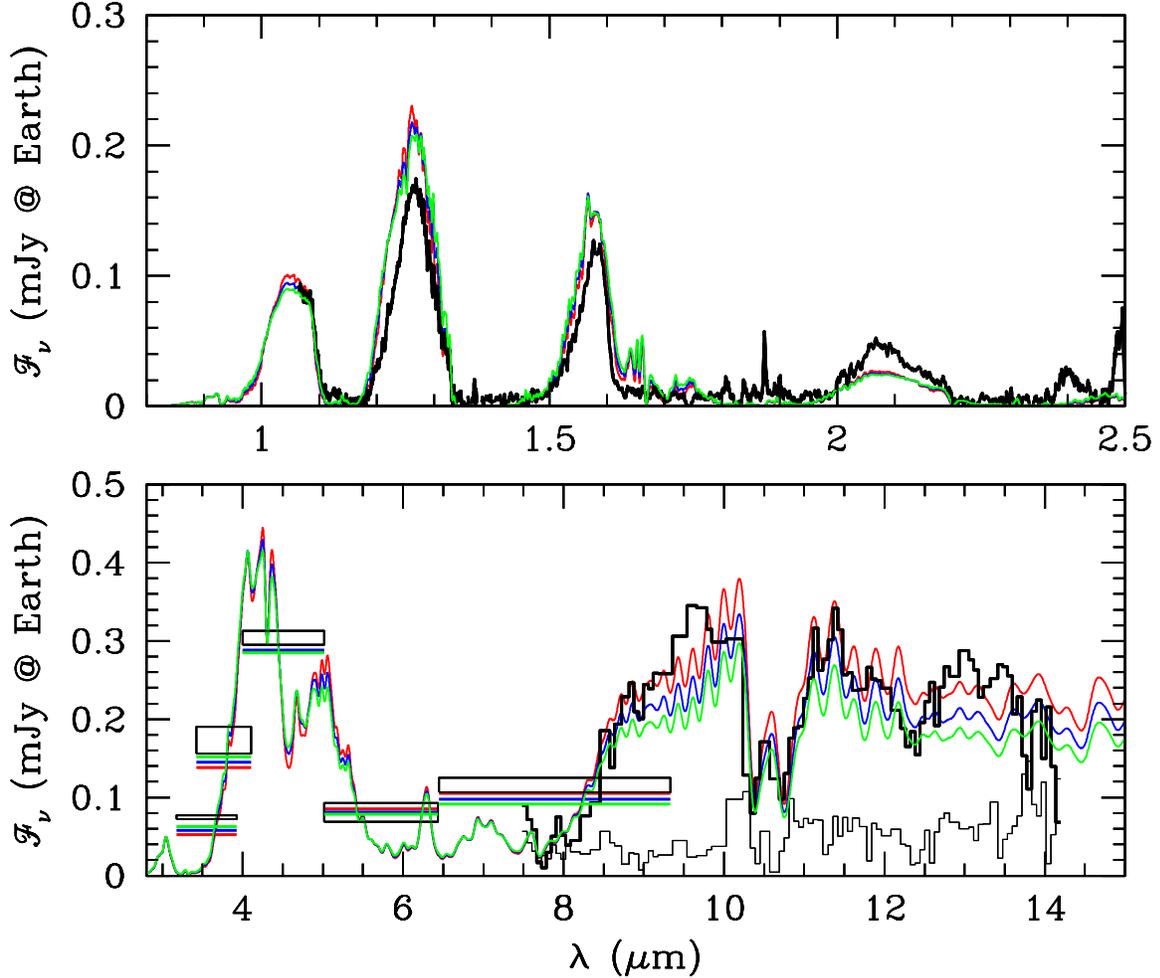}
\caption{Model spectra corresponding to the constrained solutions for
        Wolf 940 B for $\teff=575\,$K (red), 600$\,$K (blue) and 625$\,$K (green).
        See Figure 2 and Table 2.  Photometric fluxes for these models are shown 
        by horizontal lines.  All fluxes are for cloudless models with an eddy diffusion coefficient 
        of $\log K_{zz}=6$ (see text) and are absolute, based on the measured distance and the
        radii given in Table 2.  The data are shown in black, with photometric measurements shown
        as boxes with a height of $\pm 1\sigma$.  For clarity, the near-infrared spectrum (B09,
        upper panel) has been smoothed with a running 7-pixel average.  The IRS noise spectrum is
        shown in the lower panel.
        }
\end{figure}

\begin{figure}
\includegraphics[height=.8\textheight,angle=0]{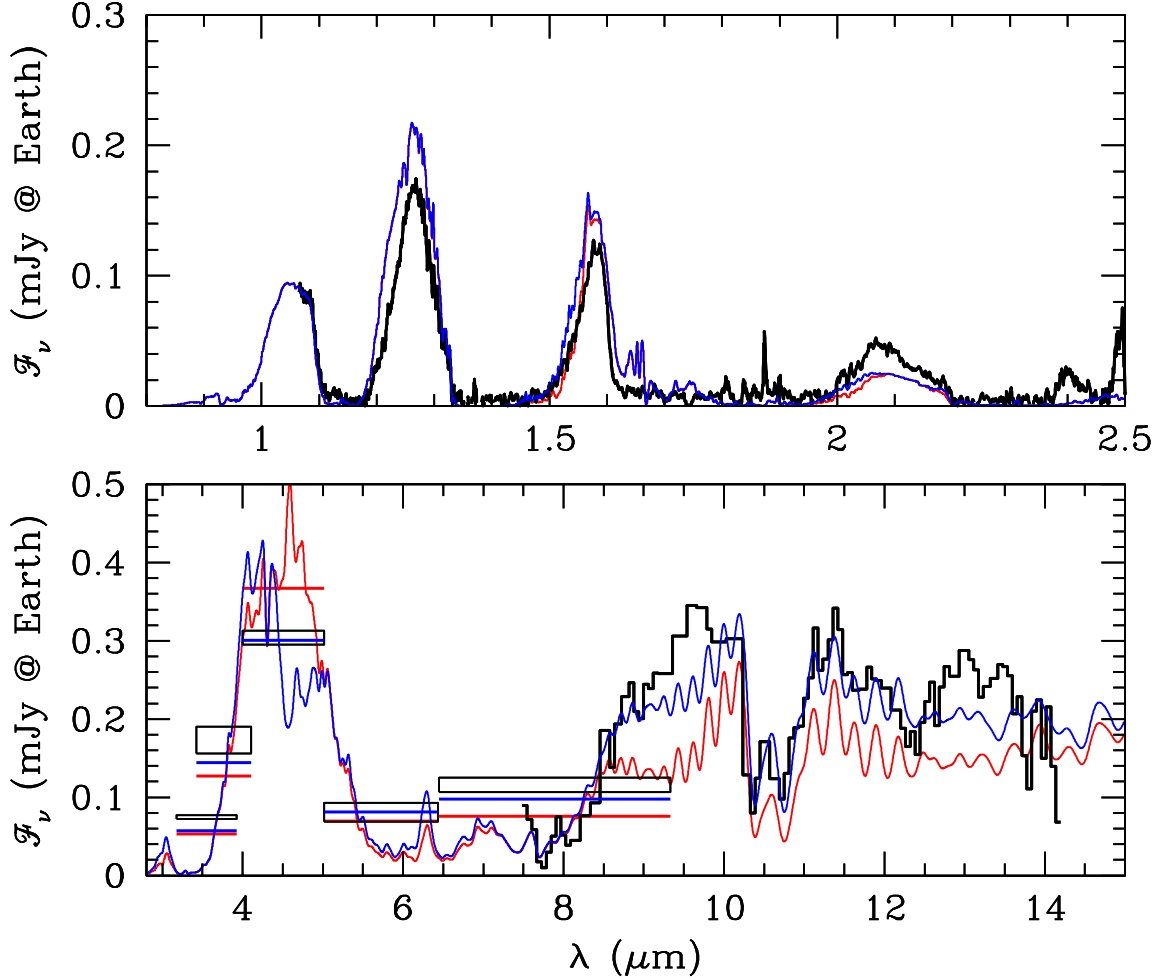}
\caption{Best fitting model spectrum with solar metallicity ($\teff=600\,$K, $\log g=4.989$, $\log K_{zz}=5.5$,
         blue curve) and the same spectrum computed with chemical equilibrium abundances ($K_{zz}=0$,
         red curve).  The large non-equilibrium abundance of CO is responsible for the strong band 
         centered at 4.66$\,\mu$m. The reduced, non-equilibrium abundance of NH$_3$ results in a higher
         flux for $\lambda \wig>8.5\,\mu$m.  See Figure 3 for additional details.
        }
\end{figure}

\clearpage 

\begin{figure}
\includegraphics[height=.7\textheight]{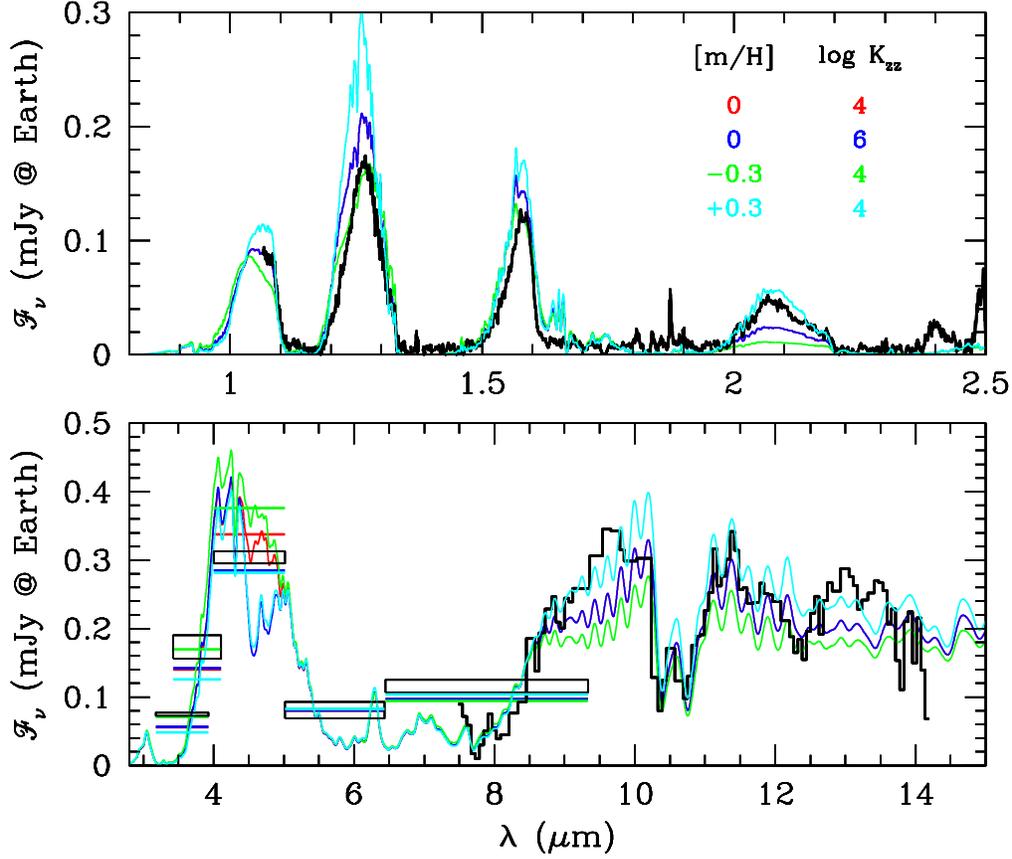}
\caption{Effect of variations in the metallicity and eddy diffusion coefficient $K_{\rm zz}$ around the
         parameters of best fitting spectrum shown in Figure 4.  All spectra shown have $\teff=600\,$K and 
         $\log g=5$.  The observed spectrum of Wolf 940 B is shown in black.  Colored curves show
         spectra computed with the metallicity and eddy diffusion coefficients shown in the legend.
         The red and blue curves overlap everywhere except in the CO band centered at 4.6$\,\mu$m.
         The parameters of the model shown in blue are almost identical to those of the best fitting
         spectrum of Figure 4.
        }
\end{figure}

\clearpage

%%\begin{figure}
%%\plottwo{f2a.eps}{f2b.eps}
%%\caption{This is an example of a multipart figure with a long figure caption 
%%that must be set as a paragraph.  The processor has to buffer the text of the
%%caption, so it is good not to be too wordy, but that would make for
%%poor communication as well.\label{fig2}}
%%\end{figure}

%% Tables should be submitted one per page, so put a \clearpage before
%% each one.

%% Two options are available to the author for producing tables:  the
%% deluxetable environment provided by the AASTeX package or the LaTeX
%% table environment.  Use of deluxetable is preferred.
%%

%% Three table samples follow, two marked up in the deluxetable environment,
%% one marked up as a LaTeX table.

%% In this first example, note that the \tabletypesize{}
%% command has been used to reduce the font size of the table.
%% Note also that the \label command needs to be placed 
%% inside the \tablecaption.

\clearpage

\begin{deluxetable}{lrr}
%\tabletypesize{\scriptsize}
\tablecaption{Astrometric and Photometric Data for the Wolf 940 System}
\tablewidth{0pt}
\tablehead{
\colhead{Property} & \colhead{Primary}  & \colhead{Secondary} }
\startdata
RA 2000, hhmmss.ss & 21 46 40.47 & 21 46 38.41 \\
Dec 2000, ddmmss.s & -00 10 25.4 & -00 10 34.6 \\
Parallax mas & 79.8$\pm$4.5\tablenotemark{a} & \nodata \\
PM RA mas yr$^{-1}$ & 765$\pm$2\tablenotemark{a} & \nodata \\
PM Dec mas yr$^{-1}$ & -497$\pm$2\tablenotemark{a} & \nodata \\
Radial Vel. kms$^{-1}$ & -31.6$\pm$12.2\tablenotemark{b} & \nodata \\
Binary separation AU  & 400$\pm$25\tablenotemark{c} & \nodata \\
Binary PA deg. & 253$\pm$1\tablenotemark{c} & \nodata \\
$V$ & 12.71$\pm$0.06\tablenotemark{d} &\nodata \\
$Y$ MKO & \nodata & 18.97$\pm$0.03\tablenotemark{c}\\
$J$ 2MASS/MKO & 8.36$\pm$0.02\tablenotemark{e} & 18.16$\pm$0.02\tablenotemark{c}\\
$H$ 2MASS/MKO & 7.83$\pm$0.03\tablenotemark{e} & 18.77$\pm$0.03\tablenotemark{c}\\
$Ks$/$K$ 2MASS/MKO & 7.49$\pm$0.03\tablenotemark{e} & 18.85$\pm$0.05\tablenotemark{c}\\
$L^{\prime}$ MKO & \nodata & 15.38$\pm$0.11\tablenotemark{c}\\
3.6 IRAC & \nodata & 16.44$\pm$0.02\tablenotemark{f}\\
4.5 IRAC & \nodata & 14.43$\pm$0.01\tablenotemark{f}\\
5.8 IRAC & \nodata & 15.38$\pm$0.15\tablenotemark{f}\\
8.0 IRAC & \nodata & 14.36$\pm$0.08\tablenotemark{f}\\
\enddata

%% Text for table notes should follow after the \enddata but before
%% the \end{deluxetable}. Make sure there is at least one \tablenotemark
%% in the table for each \tablenotetext.

\tablenotetext{a}{Harrington \& Dahn 1980}
\tablenotetext{b}{Dawson \& De Robertis 2005}
\tablenotetext{c}{Burningham et al. 2009}
\tablenotetext{d}{Mermilliod 1997}
\tablenotetext{e}{2MASS All-Sky Catalog of Point Sources, Cutri et al. 2003}
\tablenotetext{f}{Leggett et al. 2010}

%%\tablecomments{Occasionally, authors wish to append a short
%%paragraph of explanatory notes that pertain to the entire table, but
%%which are different than the caption.  Such notes should be placed in
%%a {\tt tablecomments} command like this.}

\end{deluxetable}

\clearpage

\begin{deluxetable}{cccccc}
\tablecaption{Constrained Solutions to Wolf 940 B\tablenotemark{a}}
\tablewidth{0pt}
\tablehead{
\colhead{$T_{\rm eff}$} & \colhead{$\log \ g$}  & \colhead{$\log \ L/L_{\odot}$} & 
\colhead{Mass} & \colhead{Radius} & \colhead{Age}\\
\colhead{(K)} & \colhead{(cm/s$^2$)}  & \colhead{} & 
\colhead{(M$_{\rm Jupiter)}$} & \colhead{(R$_{\odot}$)} & \colhead{(Gyr)}
}
\startdata
 550 &       4.395  & $-5.994$  &   12  &    0.1107  &   0.4 \\
 575 &       4.722  & $-6.004$  &   20  &    0.1004  &   1.0 \\
 600 &       4.989  & $-6.009$  &   31  &    0.0915  &   5.5 \\
 625  &      5.221  & $-6.012$  &   45  &    0.0839  &    10 \\
\enddata
\tablenotetext{a}{Assuming [m/H]$=$0.}
\end{deluxetable}

\clearpage

\begin{deluxetable}{rrrrrrr}
\tablecaption{Final Adopted Parameter Set for Wolf 940 B\tablenotemark{a}}
\tablewidth{0pt}
\tablehead{
\colhead{$T_{\rm eff}$} & \colhead{$\log \ g$}  & \colhead{[m/H]} &
\colhead{ $K_{zz}$} &
\colhead{Mass} & \colhead{Radius} & \colhead{Age}\\
\colhead{K} & \colhead{}  & \colhead{} & \colhead{cm$^2$s$^{-1}$} & 
\colhead{M$_{\rm Jupiter}$} & \colhead{R$_{\odot}$} & \colhead{Gyr}
}
\startdata
 585 -- 625 &  4.83 -- 5.22  &  0.24 $\pm$ 0.09  &  $10^4$--$10^6$ & 24 -- 45  &    
0.097 -- 0.084  &   3 -- 10 \\
\enddata
\tablenotetext{a}{Temperature, gravity, mass and radius are correlated and are listed in order of increasing age. Metallicity has been determined by application of the Johnson \& Apps (2008) $M_K$:$V-K$ relationship to Wolf 940 A. Age is constrained by the kinematic and H$\alpha$ properties of Wolf 940 A.}
\end{deluxetable}

\end{document}